\documentstyle[amssymb,psfig]{mn2e}

\title[Age and mass distribution of the LMC star cluster system]{How
well do we know the age and mass distributions of the star cluster
system in the Large Magellanic Cloud?}

\author[Richard de Grijs and Peter Anders]{Richard de
Grijs$^1$\thanks{E-mail: R.deGrijs@sheffield.ac.uk} and Peter Anders$^2$\\
$^1$ Department of Physics \& Astronomy, The University of Sheffield,
Hicks Building, Hounsfield Road, Sheffield S3 7RH\\ 
$^2$ Institut f\"ur Astrophysik, Georg-August-Universit\"at,
Friedrich-Hund-Platz 1, 37077 G\"ottingen, Germany 
}

\date{Received date; accepted date}
\pubyear{2005}

\begin{document}
\maketitle

\begin{abstract}
The Large Magellanic Cloud (LMC) star cluster system offers the unique
opportunity to independently check the accuracy of age (and the
corresponding mass) determinations based on a number of complementary
techniques. Using our sophisticated tool for star cluster analysis
based on broad-band spectral energy distributions (SEDs),
``AnalySED'', we reanalyse the Hunter et al. (2003) LMC cluster
photometry. Our main aim is to set the tightest limits yet on the
accuracy of {\it absolute} age determinations based on broad-band
SEDs, and therefore on the usefulness of such an approach. Our
broad-band SED fits yield reliable ages, with statistical absolute
uncertainties within $\Delta\log( \mbox{Age/yr}) \simeq 0.4$
overall. The systematic differences we find with respect to previous
age determinations are caused by conversions of the observational
photometry to a different filter system, thus leading to
systematically inaccurate results.

The LMC's cluster formation rate (CFR) has been roughly constant
outside of the well-known age gap between $\sim 3$ and 13 Gyr, when
the CFR was a factor of $\sim 5$ lower. Using a simple approach to
derive the characteristic cluster disruption time-scale, we find that
$\log(t_4^{\rm dis}/{\rm yr}) = 9.9 \pm 0.1$, where $t_{\rm dis} =
t_4^{\rm dis} (M_{\rm cl}/10^4 {\rm M}_\odot)^{0.62}$. This long
characteristic disruption time-scale implies that we are observing the
{\it initial} cluster mass function (CMF). We conclude that while the
older cluster (sub)samples show CMF slopes that are fully consistent
with the $\alpha \simeq -2$ slopes generally observed in young star
cluster systems, the youngest mass and luminosity-limited LMC cluster
subsets show shallower slopes (at least below masses of a few $\times
10^3$ M$_\odot$), which is contrary to dynamical expectations. This
may imply that the initial CMF slope of the LMC cluster system as a
whole is {\it not} well represented by a power-law, although we cannot
disentangle the unbound from the bound clusters at the youngest ages.
\end{abstract}


\section{Introduction}
\label{intro.sec}

The debate regarding the true underlying mass function of young star
cluster systems is presently very much alive, because it bears on the
very essence of the star-forming process, as well as on the formation,
assembly history and evolution of the clusters' host galaxies over
cosmic time. Although observational evidence of young cluster systems
in many star cluster-forming interacting and starburst galaxies
appears to indicate that they are well represented by power-law
cluster {\it luminosity} functions (CLFs; see de Grijs et al. 2003d
for a recent comprehensive review), rapid changes to the stellar
population properties, combined with a (possibly significant) age
range within a given cluster system may conspire to give rise to a
power-law-like CLF, even if the true underlying cluster {\it mass}
function (CMF) is not a power law (Meurer 1995; Miller et al. 1997;
Fritze--v. Alvensleben 1998, 1999; de Grijs et al. 2001, 2003a,b;
Hunter et al. 2003, hereafter H03). It is, therefore, obviously very
important to age date the individual clusters and to correct the
observed CLF to a common age, before interpreting the cluster
luminosities in terms of the corresponding mass distribution
(Fritze--v. Alvensleben 1999; de Grijs et al. 2001, 2003a,b;
H03). This is particularly important if the age spread within a
cluster system is a significant fraction of the system's mean age.

Even with the high spatial resolution offered by the advanced suite of
instruments onboard the {\sl Hubble Space Telescope}, extragalactic
star clusters at distances beyond the Local Group can only be studied
through their integrated properties. However, we {\it do} have access
to a statistically significant young to intermediate-age {\it
resolved} star cluster system on our doorstep, notably in the Large
Magellanic Cloud (LMC). By combining {\it integrated} properties with
{\it resolved} stellar population studies, the LMC cluster system
offers the unique chance to independently check the accuracy of age
(and corresponding mass) determinations based on broad-band spectral
energy distributions (SEDs). This is what we set out to do in this
paper, based on the largest, most homogeneous and complete set of
integrated LMC cluster photometry presently available. With our new
cluster age estimates, spanning ages from a few Myr up to 10 Gyr,
determined in an internally consistent fashion and with a firm handle
on the associated uncertainties, we now also have the largest,
homogeneous sample of cluster masses with well-defined uncertainties
to date, which we use to explore the evolution of the CMF.

In Section \ref{data.sec} we discuss the details of the LMC cluster
sample we base our analysis on, and explore the uncertainties
associated with the age and mass determinations. Section
\ref{formdis.sec} discusses the formation and disruption history of
the LMC star cluster sample. We use this as our basis for the
interpretation of the LMC cluster mass distribution in terms of the
initial and evolved distributions of cluster masses in Section
\ref{cmf.sec}. Finally, in Section \ref{summary.sec} we summarise our
results and conclusions.

\section{Data and accuracy analysis}
\label{data.sec}

\begin{figure*}
\hspace*{0.7cm}
\psfig{figure=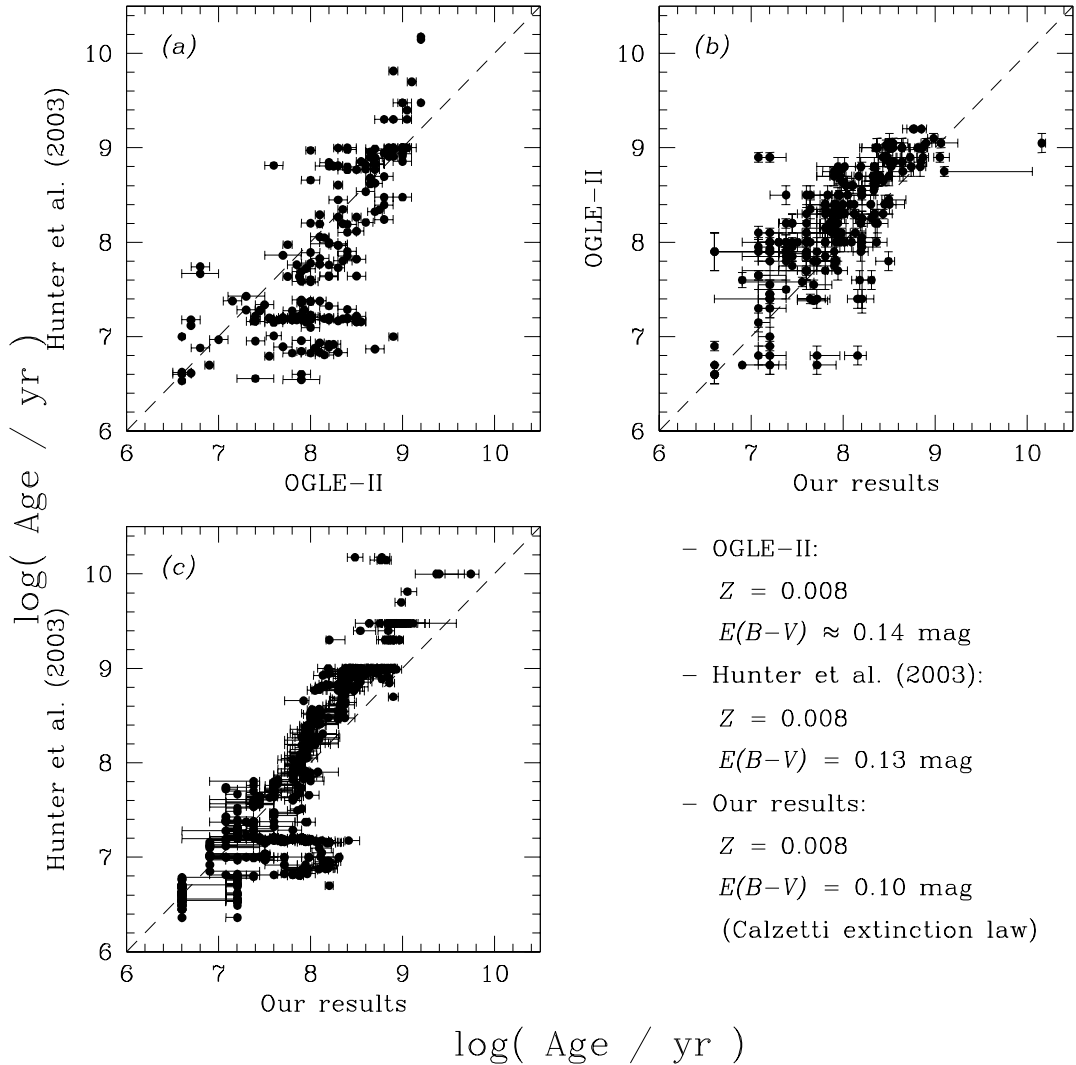,width=15cm}
\caption{\label{lmccf.fig}Comparison of age determination for LMC
clusters in common between Hunter et al. (2003), the OGLE-II team of
Pietrzy\'nski \& Udalski (2000), and this paper. The assumptions on
the overall metallicity and extinction properties of the cluster
sample are indicated in the legend.}
\end{figure*}

The basis for our detailed comparison of the age estimates for the LMC
star cluster system is provided by (i) the $UBVR$ broad-band SEDs of
H03, based on Massey's (2002) CCD survey of the Magellanic Clouds, and
(ii) the OGLE-II data set (Udalski, Kubiak \& Szyma\'nski 1997), and
in particular their homogeneously determined ages for some 600
clusters, using colour-magnitude diagrams (CMDs; Pietrzy\'nski \&
Udalski 2000).

H03 determined the ages for the individual LMC clusters by comparing
their broad-band SEDs with a variety of cluster evolutionary
models. Specifically, they used the Starburst99 (Leitherer et
al. 1999) with $Z = 0.008$ for ages up to 1 Gyr, while for older ages
they used a combination of the $UBV$ colours of Searle, Sargent \&
Bagnuolo (1973), the $(V-R)_{\rm C}$ colours for globular clusters
from Reed (1985), and the Charlot \& Bruzual (1991) simple stellar
population (SSP) models for the evolution of the $V$-band luminosity
in the age range from 1 to 10 Gyr. They then used a combination of two
colour-colour diagrams, $(U-B)$ vs. $(B-V)$ and $(B-V)$ vs. $(V-R)$,
to obtain their age estimates, by comparing the cluster positions to
the model tracks in those diagrams.

The colours and magnitudes of their sample clusters were corrected for
reddening using a blanket reddening of $E(B-V) = 0.13$ mag for all
clusters, where they adopted the extinction curve of Cardelli et
al. [1989; $R_V \equiv A_V/E(B-V)= 3.10$].

The Pietrzy\'nski \& Udalski (2000) ages are based on detailed
isochrone fits to the OGLE-II colour-magnitude data, using the Padova
isochrones for $Z=0.008$. They dereddened their photometry using the
extinction values for 84 LMC subfields determined by Udalski et
al. (1999), $E(B-V) = 0.143 \pm 0.020$ mag. These extinction
values are based on Schlegel et al.'s (1998) reddening maps for the
Galactic foreground extinction, on average $E(B-V)=0.075$ mag, and
their $R_V = 3.24$.

Finally, here we reanalyse the Massey (2002) data set, also assuming
$Z = 0.008$, for comparison purposes (although our analysis tool
allows us to leave the metallicity as a free parameter; see Section
\ref{degeneracies.sec}). For the total extinction towards the LMC
clusters, we assumed $E(B-V) = 0.10$ mag, using the Calzetti
attenuation law (Calzetti 1997, 2001; Calzetti et al. 2000; Leitherer
et al. 2002) with $R_V = 4.05$. This corresponds to $E(B-V) \simeq
0.13$ mag for both the Cardelli et al. (1989) and the Schlegel et
al. (1998) extinction laws.

In a series of recent papers, we developed a sophisticated tool for
star cluster analysis based on broad-band SEDs, ``AnalySED'', which we
tested extensively both internally (de Grijs et al. 2003c,d; Anders et
al. 2004) and externally (de Grijs et al. 2005), using both
theoretical and observed young to intermediate-age ($t \lesssim 3
\times 10^9$ yr) star cluster SEDs, and the {\sc galev} SSP models
(Kurth et al. 1999; Schulz et al. 2002). We increased the accuracy for
younger ages by the inclusion of an extensive set of nebular emission
lines, as well as gaseous continuum emission (Anders \& Fritze-v.
Alvensleben 2003). We concluded that the {\it relative} ages within a
given cluster system can be determined to a very high accuracy
depending on the specific combination of passbands used (Anders et
al. 2004). Even when comparing the results of different groups using
the same data set, we can retrieve prominent features in the cluster
age distribution to within $\sigma_t \equiv \Delta \langle \log( {\rm
Age / yr} ) \rangle \le 0.35$ (de Grijs et al. 2005), which confirms
that we understand the uncertainties associated with the use of our
AnalySED tool to a very high degree.

The aim of this section is to compare our new age determinations with
(i) those of H03 using the same data set but a different approach, and
(ii) those of Pietrzy\'nski \& Udalski (2000) using the independent
(and presumably more accurate) method of CMD fitting. This will allow
us to set the tightest limits yet on the accuracy of {\it absolute}
age determinations based on broad-band SEDs, and therefore on the
usefulness of such an approach in general.

First, we compare the independently determined ages of the clusters in
common between the H03 and OGLE-II results. Fig. \ref{lmccf.fig}a
shows the distribution of the data points in the relevant parameter
space, where the dashed line represents equality between both age
estimates for a given cluster. It is immediately clear that there is a
systematic difference between the H03 and OGLE-II ages, the magnitude
of which is a clear function of age. The effect is indeed significant;
a straightforward linear regression results in a slope that is
significantly different from unity, $1.47 \pm 0.07$. We will explore
the origin of this systematic effect in Section \ref{filters.sec}
below. It is also clear that, in units of $\log(\mbox{Age/yr})$, the
intrinsic scatter in the data points is somewhat greater at younger
ages; this is merely an effect of the poorer intrinsic age accuracy
for older ages imposed by the models.

Secondly, we compare our redetermined ages to both those of OGLE-II
and H03, in Figs. \ref{lmccf.fig}b and c, respectively. We point out
that because of the limiting $V$-band magnitude of the OGLE-II data,
their age estimates are only reliable for ages $\lesssim 1$ Gyr.
Consequently, the comparison between the OGLE-II CMD-based ages and
our age determinations in this paper are only valid for ages $\lesssim
1$ Gyr, as is also evident from the lack of data points for $\log({\rm
Age/yr}) \gtrsim 9.0$ in Fig. \ref{lmccf.fig}b. For the older sample
clusters, for which age estimates are available based on spectral
features or CMD analysis, the comparison is extended on a one-by-one
basis in Section \ref{other.sec}.

The systematic effect shown by the H03 data seems to have disappered
when we compare our results to those of the OGLE-II team; the
resulting slope is $1.05 \pm 0.05$, with an x-axis intercept at $\log(
\mbox{Age/yr} ) = -0.11 \pm 0.39$. This excellent agreement indicates
that our assumption that the overall extinction is well represented by
$E(B-V) = 0.10$ mag (assuming the Calzetti attenuation law) is well
justified; a change in extinction of $|\Delta E(B-V)|=0.05$ would
correspond to an offset from the line of equality of $|\Delta\log(
\mbox{Age/yr} )| \simeq 0.10$. This effect is a direct result of the
age-extinction, and possibly (to a lesser extent) also of the
age-metallicity degeneracy (see Section \ref{degeneracies.sec}). The
systematic effect with respect to the H03 results is still visible,
although the scatter in the results is -- perhaps not surprisingly
because we used the same data set -- much smaller than when we compare
our results to those of OGLE-II. Table \ref{regression.tab} contains
the full numerical details for a quantitative comparison among the age
determinations based on the samples discussed here. We note that any
remaining offset between our results and those of the OGLE-II team,
despite having used the same assumptions for the total extinction
towards the clusters, may have been introduced by the exact value for
the distance modulus assumed by Pietrzy\'nski \& Udalski (2000), $m-M
= 18.23$ mag, which is an essential ingredient for age determinations
based on CMD fitting. If they had assumed a nominal $m-M = 18.50$ mag,
their resulting age estimates would have been younger by $\Delta
\log({\rm Age / yr}) \sim 0.2-0.4$, depending on the age range
considered (larger offsets result for younger ages).

\begin{table*}
\caption[ ]{\label{regression.tab}Quantitative comparison among age
determinations.}
{\scriptsize
\center{
\begin{tabular}{llcrcc}
\hline
\hline
\multicolumn{1}{c}{Source 1} & \multicolumn{1}{c}{Source 2} &
\multicolumn{1}{c}{Slope} & \multicolumn{1}{c}{Intercept} &
\multicolumn{1}{c}{Scatter, $\langle\sigma_t\rangle$} &
\multicolumn{1}{c}{Correlation} \\
& & & \multicolumn{1}{c}{[$\log(\mbox{Age/yr})$]} &
\multicolumn{1}{c}{[$\langle\Delta\log(\mbox{Age/yr})\rangle$]} &
\multicolumn{1}{c}{coefficient, $\mathcal{R}$}\\
\hline
OGLE-II    & H03             & $1.44 \pm 0.07$ & $-3.85 \pm 0.61$ & 0.62 & 0.74 \\
OGLE-II    & H03 (corrected) & $1.04 \pm 0.05$ & $-0.47 \pm 0.44$ & 0.44 & 0.74 \\
This paper & OGLE-II         & $1.05 \pm 0.05$ & $-0.11 \pm 0.39$ & 0.42 & 0.74 \\
This paper & H03             & $1.38 \pm 0.03$ & $-2.96 \pm 0.20$ & 0.47 & 0.88 \\
This paper & H03 (corrected) & $0.98 \pm 0.02$ & $ 0.23 \pm 0.14$ & 0.33 & 0.88 \\
\hline
\end{tabular}
}}
\end{table*}

\subsection{Filter transmission curves}
\label{filters.sec}

\begin{figure*}
\hspace*{0.7cm}
\psfig{figure=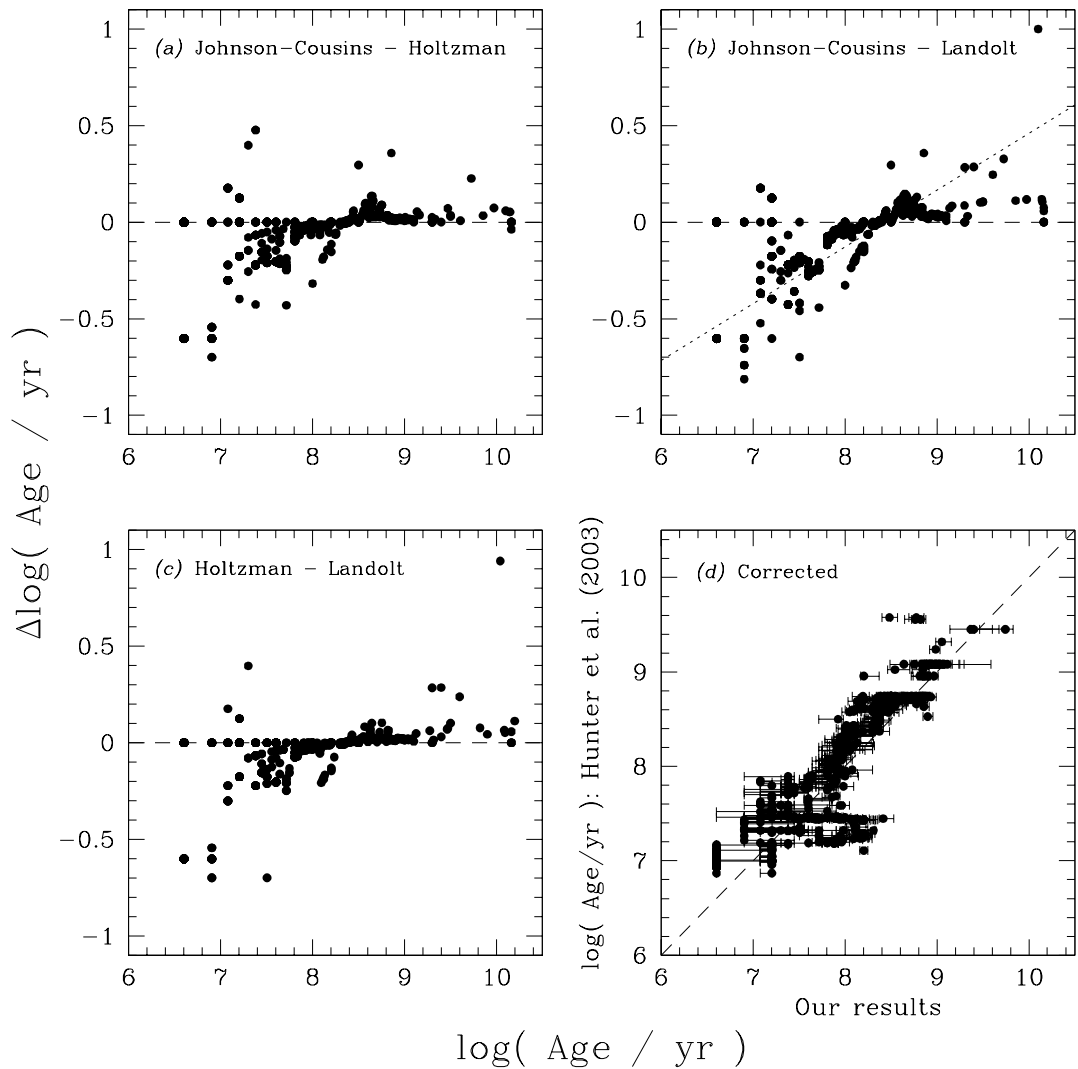,width=15cm}
\caption{\label{cffilters.fig}(a)--(c) The effects of colour
transformations between filter systems on the derived ages for the
Massey (2002) LMC cluster photometry, assuming $Z=0.008$ and
$E(B-V)=0.10$ mag. We have chosen to omit the formal uncertainties on
the data points for reasons of clarity. (d) Comparison of the
corrected age determinations for the H03 sample to our
redeterminations in this paper.}
\end{figure*}

In the previous section we noted a significant systematic effect
between the age differences of H03 on the one hand, and those of both
the OGLE-II team and our own redeterminations on the other. H03
converted the Starburst99 Johnson $(V-R)$ colour to the Cousins
system. As we showed in de Grijs et al. (2005; their figs. 10 and 11),
filter conversions may be responsible for significant differences in
the resulting age determinations. In order to investigate the
possibility of this effect playing a role for these data, we carefully
analysed the properties of the original and converted systems
used. Massey (2002) calibrated his photometry using the Landolt (1992)
standard stars, so that his photometry should be analysed using the
filter set used to obtain the standard-star photometry. The Landolt
(1992) filter curves are very close to the filter transmission curves
used by Holtzman et al. (1995), who also kindly supplied us with the
original, unpublished Landolt KPNO curves (J. Holtzman, priv. comm.).
For completeness, we will therefore assess the differences in the
resulting ages for our LMC cluster sample based on using the
``standard'' Johnson-Cousins system (as done by H03), the curves used
by Holtzman et al. (1995), and the original KPNO curves used by
Landolt (1992).

The results of this are shown in Fig. \ref{cffilters.fig}. Fig.
\ref{cffilters.fig}c shows that the filter transmission curves used by
Holtzman et al. (1995) and Landolt (1992), respectively, indeed yield
relatively similar age estimates. The most important comparison figure
is displayed in panel b, however. The systematic trend seen here
mimics that seen in Figs. \ref{lmccf.fig}a and c, in the sense that if
one uses the Johnson-Cousins filter system (even if based on the
appropriate conversion equations) instead of the native Landolt KPNO
system, one will obtain lower ages than expected at the young-age end
of the age range covered by our LMC cluster sample, and vice versa. To
provide further evidence for this scenario, we applied a simple linear
regression to the data points in Fig. \ref{cffilters.fig}c (shown by
the dotted line), and applied the resulting equation as a first-order
correction to H03's age determinations in both Figs. \ref{lmccf.fig}a
and c. The resulting, corrected age comparison is shown in
Fig. \ref{cffilters.fig}d, while the numerical values are once again
included in Table \ref{regression.tab}. It thus appears that the
systematic differences in LMC cluster ages between the H03 results and
those of OGLE-II and ourselves are indeed caused by their conversions
of the photometry to a different filter system.

\subsection{The age-metallicity and age-extinction degeneracies}
\label{degeneracies.sec}

\begin{figure*}
\hspace*{0.7cm}
\psfig{figure=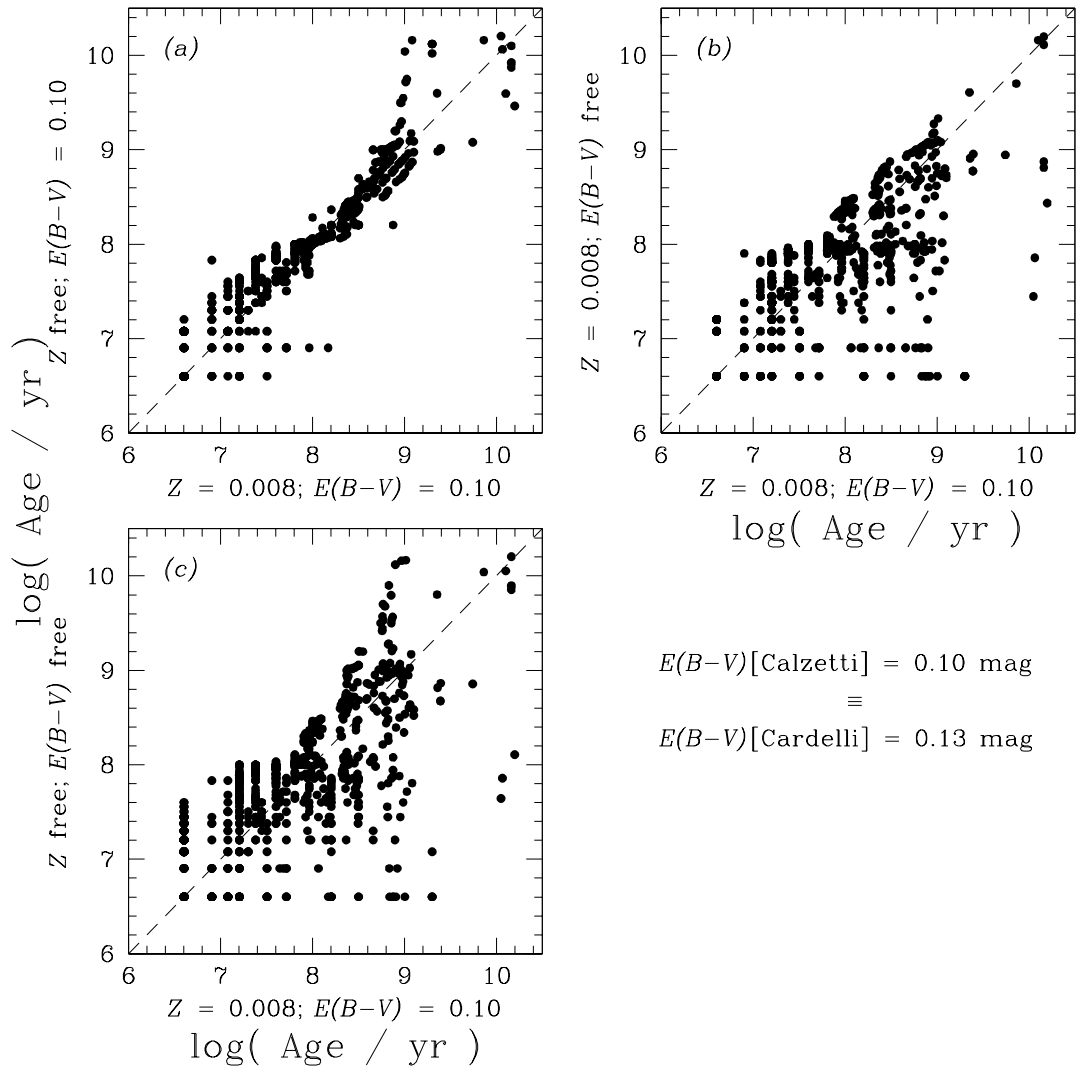,width=15cm}
\caption{\label{internal.fig}The effects of (a) the age-metallicity,
(b) the age-extinction, and (c) the combined degeneracies. Again,
error bars have been omitted for reasons of clarity.}
\end{figure*}

In the discussion of our results, we have thus far assumed a fixed
metallicity of $Z=0.008$ and a fixed extinction of $E(B-V)=0.10$ mag
(assuming the Calzetti attenuation law) for all of our sample
clusters. The AnalySED tool has been developed to also provide
independent information on a cluster's metallicity and extinction,
provided that a significant number of data points defining the SED are
available to match the number of free parameters (cf. de Grijs et
al. 2003c,d; Anders et al. 2004).

Therefore, we have also reanalysed the LMC cluster photometry assuming
(i) variable metallicity and $E(B-V)=0.10$ mag, (ii) $Z=0.008$ and
variable extinction, and (iii) variable extinction and metallicity,
the results of which are shown in Figs. \ref{internal.fig}a, b and c,
respectively. From close scrutiny of these panels, it appears that the
age-metallicity degeneracy is least important {\it for this cluster
sample}; this implies that the assumption of $Z=0.008$ adopted by H03,
OGLE-II and ourselves in the previous sections is a reasonable
approximation of the average LMC cluster metallicity. We also note
that adopting $Z=0.008$ as a reasonable approximation for the average
cluster metallicity is supported -- at least for clusters younger than
$\sim 3$ Gyr -- by a number of spectroscopic studies based on
individual cluster stars (see, e.g., Olszewski et al. 1991, their
fig. 11).

The effects of the age-extinction degeneracy are dramatic for this
sample, however. On average, we find that if we had not assumed a
fixed extinction value, the resulting ages would have been smaller
(although there remains significant scatter about the dashed line of
equality).  However, we caution that some of this scatter is likely
artificial; as we showed in de Grijs et al. (2003d), the
age-extinction degeneracy is artificially worsened (in the sense that
clusters are artificially assigned younger ages) if the wavelength
baseline used for the multi-parameter cluster SED analysis lacks
wavelength coverage longward of the $R$ filter, as is the case
here. In essence, this is shown by the enhanced population of the two
youngest age bins (at 4 and 8 Myr) in Figs. \ref{internal.fig}b and c
for the models without restrictions on $E(B-V)$. This situation can be
avoided successfully by adopting a fixed value for either the cluster
population's (average) extinction or metallicity (de Grijs et
al. 2003d).

Thus, in view of these considerations, this implies that our results
are indeed very comparable to those obtained from the OGLE-II data.

\subsection{Other comparisons}
\label{other.sec}

\begin{figure*}
\hspace*{0.7cm} 
\psfig{figure=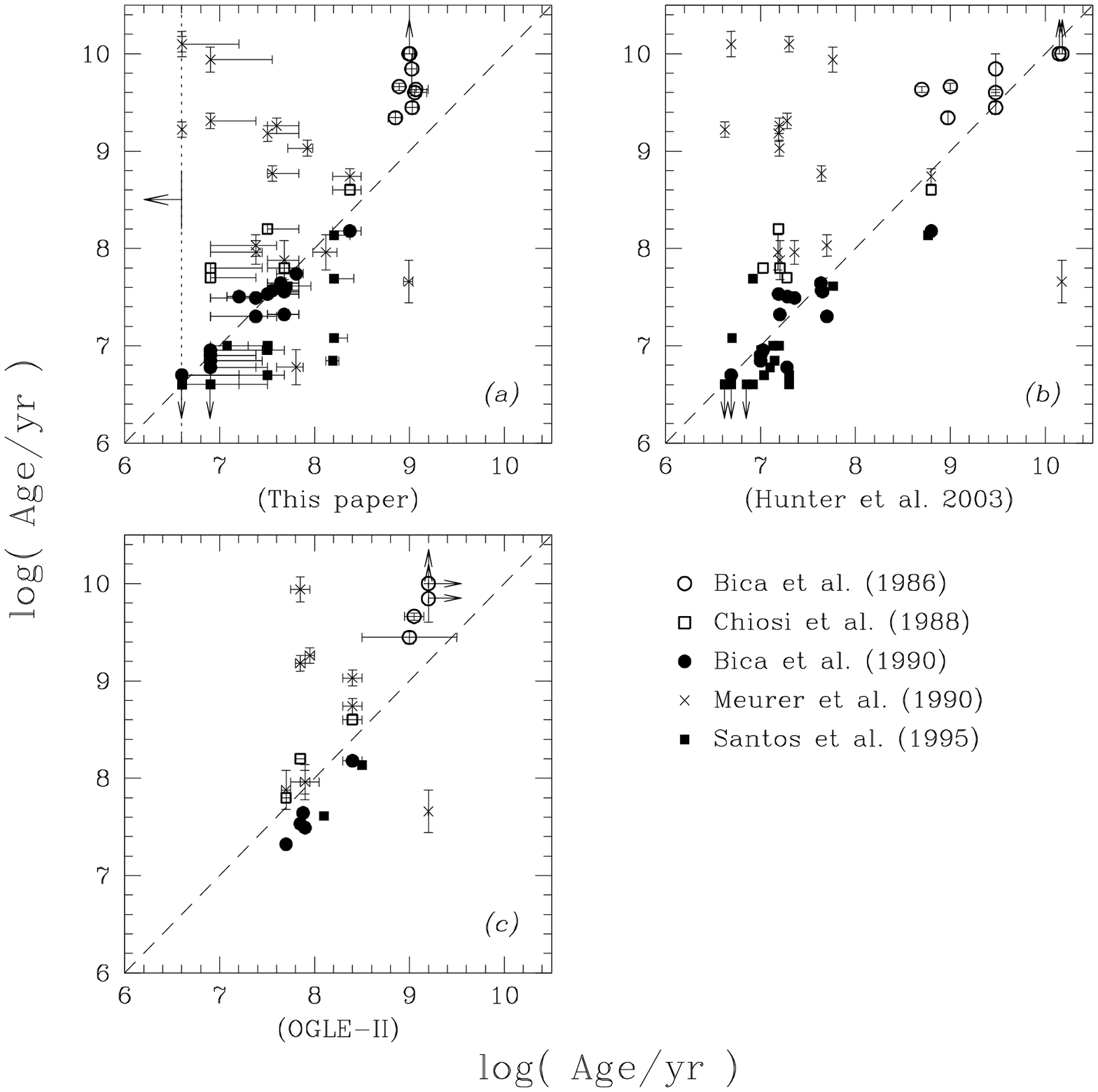,width=15cm}
\caption{\label{colours.fig}Comparison with other age determinations,
based on broad-band photometry.}
\end{figure*}

Finally, to assess the robustness of the individual ages, we compare
the LMC cluster age determinations of H03 (uncorrected), OGLE-II and
ourselves with those obtained for smaller cluster samples using a
variety of techniques. In Fig. \ref{colours.fig} we have included
those samples containing $\ge 5$ clusters in common with our sample
for which the authors determined their ages based on broad-band
photometry. With the exception of the Meurer et al. (1990) results,
who used the ultraviolet colours $C(18-31)_0$ and $C(25-31)_0$, all
other age determinations are based on blue optical colours. Bica et
al. (1986, 1990) used the $C(43-45)_0$ vs. $C(45-{\rm H}\beta)_0$
vs. SWB type (Searle, Wilkinson \& Bagnuolo 1980) diagnostics, while
Chiosi et al. (1988; $UBV$ clours) and Santos et al. (1995; $(U-B)_0$
colours) limited themselves to the ``standard'' set of broad-band
filters.

Although the Meurer et al. (1990) ages are clearly systematically
greater than any of the ages determined by H03, OGLE-II or us, the
other age determinations are statistically similar to those discussed
elsewhere in this paper. This is encouraging regarding the robustness
of our results; the relatively large scatter is a side effect of the
poor age resolution for broad-band colours only. The fact that Meurer
et al.'s (1990) ultraviolet-based age determinations are clearly
greater than ours underscores the difficulties involved in using this
wavelength range alone for age determinations. The main uncertainties
inherent to this at ultraviolet wavelengths are the uncertain stellar
population synthesis and uncertain extinction corrections.

\begin{figure*}
\hspace*{0.7cm}
\psfig{figure=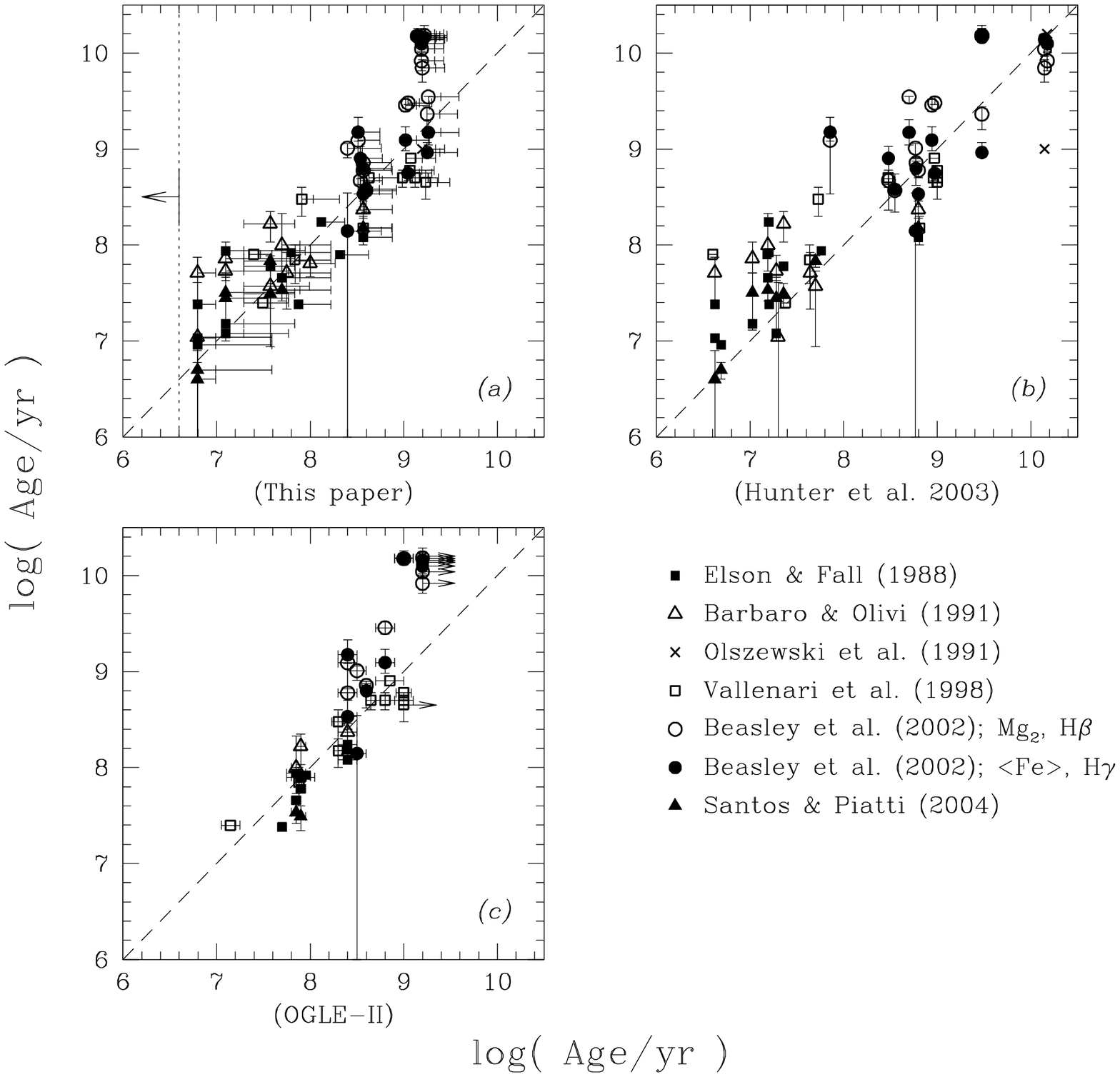,width=15cm}
\caption{\label{features.fig}Comparison with other age determinations,
based on either spectral features or CMD fitting.}
\end{figure*}

In Fig. \ref{features.fig} we show a similar comparison between the
ages discussed elsewhere in this paper, and those obtained
independently based on spectral features or CMD fits. The similarities
between our and H03's broad-band SED fits and the OGLE-II fits on the
one hand, and the determinations from the literature are striking, as
is the relatively small scatter about the (dashed) lines of
equality. Both the simplest of these age determinations, Elson \&
Fall's (1988) ages based on the clusters' main-sequence turn-off
magnitudes (assuming a common distance to their entire sample), as
well as the CMD fits by Olszewski et al. (1991) and Vallenari et
al. (1998), and the more sophisticated determinations based on
distinct spectral features by Beasley et al. (2002; see figure legend)
and Santos \& Piatti (2004; based on equivalent width measurements of
the integrated spectra) match our broad-band SED ages very well. We
also emphasize that Barbaro \& Olivi's (1991) ages based on
ultraviolet {\it spectra} return reliable ages, as opposed to those
based on the ultraviolet {\it colours} discussed in relation to
Fig. \ref{colours.fig}. H03, based on a comparison involving fewer
objects taken from the literature, also concluded that the general
match between their results and CMD-based ages was satisfactory,
although they stated that for the oldest cluster colours tend to
underestimate the ages compared to CMD fits. However, we have now
shown that most (but possibly not all; see below) of this discrepancy
was most likely caused by the systematic effects introduced by the
filter conversion they employed. The small number of clusters
available for this comparison (and the smaller number used by H03) did
not allow H03 to notice the systematic offset discussed in the
previous sections. Based on the small number of clusters aged $\sim 1$
Gyr for which comparison data is available in the literature, we
tentatively conclude that our broad-band SED analysis may have led to
underestimates of the cluster ages for $\log({\rm Age/yr}) \simeq 9$,
in a similar sense as seen by H03. This tentative conclusion is
supported by Fig. \ref{features.fig}a, where our new age
determinations appear somewhat lower than those determined from CMD
analysis or spectral features for these ages.

Based on this comparison with independently determined ages for
subsamples of LMC clusters, we conclude that our broad-band SED fits
yield reliable ages, with statistical absolute uncertainties within
$\Delta\log( \mbox{Age/yr}) \simeq 0.4$ overall, based on the scatter
about the lines of equality. Thus, in addition to our conclusion in de
Grijs et al. (2005) that we can retrieve prominent features in the
cluster age distribution to within $\Delta \langle \log( {\rm Age /
yr} ) \rangle \le 0.35$, we have now also shown that the {\it
intrinsic} statistical uncertainties involved in cluster age
determinations based on broad-band SEDs are of a very similar
magnitude.

\section{Cluster formation and disruption rates}
\label{formdis.sec}

Using the newly determined and improved age estimates for the largest
LMC cluster sample spanning the most extensive age and mass ranges to
date, we now have the means to constrain the past cluster formation
rate (CFR), as well as the effects of cluster disruption, to
unprecedented detail.

\subsection{Derivation of the characteristic cluster disruption
time-scale}

\begin{figure}
\psfig{figure=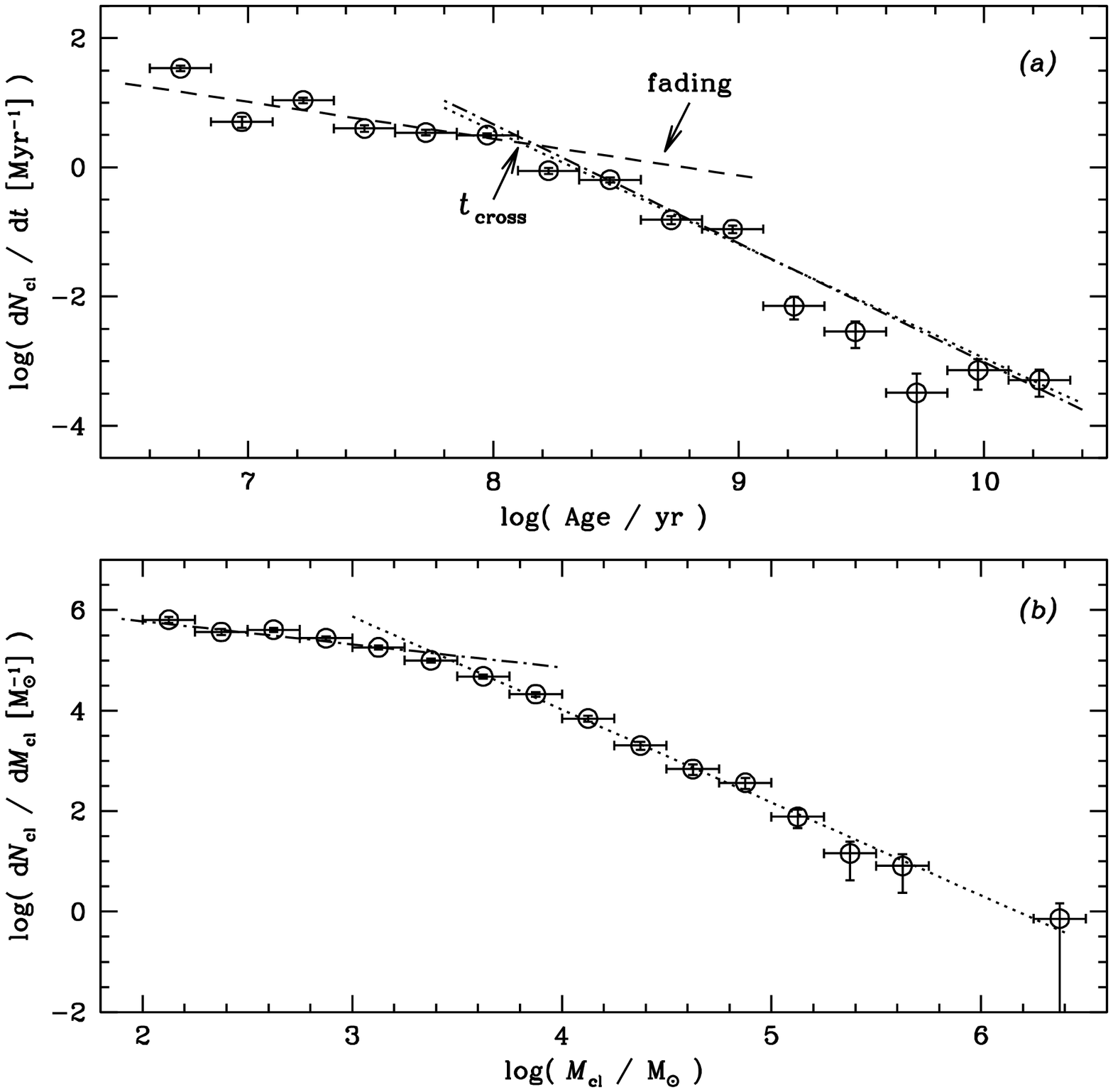,width=8.8cm}
\caption{\label{agemasshist.fig}{\it (a)} -- The LMC cluster formation
rate (in number of clusters per Myr) as a function of age. The dashed
line is the least-squares power-law fit to the fading, non-disrupted
clusters, for a constant ongoing cluster formation rate. The dotted
and dash-dotted lines are the disruption lines for the most likely age
ranges where disruption may dominate evolutionary fading, as discussed
in the text; {\it (b)} -- Mass spectrum of the LMC clusters (number of
clusters per unit mass range); the dotted line represents a power-law
fit to the CMF that is as yet unaffected by disruption (see text); the
slope of the dash-dotted line (which was shifted vertically to match
the observational data) is entirely determined by the initial CMF
slope determined from the dotted line, and stellar population
synthesis, as described in the text.}
\end{figure}

In Fig. \ref{agemasshist.fig}a we display the number of clusters
formed per unit age range. Evolutionary fading, as predicted by
stellar population synthesis models, will cause this number to slowly
decline from the youngest ages upward. This effect is shown by the
dashed line in Fig. \ref{agemasshist.fig}a, of which the slope is
entirely determined by the details of stellar population synthesis; we
have only shifted this line to best match the data points for
$\log({\rm Age/yr}) \le 8$. Both internal and external effects, such
as two-body relaxation, disk and bulge shocking, and the tidal effects
caused by the underlying galactic gravitational potential (even in the
low-density environment of the LMC; cf. Lamers, Gieles \& Portegies
Zwart 2005), leading to tidal stripping and to evaporation of a
fraction of the low-mass cluster stars, will result in the (gradual)
dissolution of star clusters. Simple (instantaneous) disruption theory
(Boutloukos \& Lamers 2003) predicts that, assuming a constant CFR and
that the characteristic cluster disruption time-scale is a function of
cluster mass, the effects of cluster disruption will dominate from a
certain age onwards, giving rise to the characteristic double
power-law seen in Fig. \ref{agemasshist.fig}a.

The location of the crossing point of the fading and disruption lines
in Fig. \ref{agemasshist.fig}a, denoted by ``$t_{\rm cross}$'', is a
key ingredient to derive the characteristic cluster disruption
time-scale for a given cluster population (Boutloukos \& Lamers
2003). The other parameters required to derive this time-scale are the
limiting magnitude, $M_V = -3.5$ mag (H03), $E(B-V)=0.10$ mag or $A_V
= 0.41$ mag (Section \ref{data.sec}), $(m-M)_{\rm LMC} = 18.48$ mag,
and the reference magnitude, $M_V^{\rm ref} = -11.3668$ mag, for a
$10^4$ M$_\odot$ cluster at an age of $10^8$ yr, which we obtained
from the {\sc galev} models for $Z = 0.4$ Z$_\odot$, assuming a
``standard'' Salpeter-like stellar IMF from 0.15 to $\sim 70$
M$_\odot$. For the crossing time we find $\log(t_{\rm cross}/{\rm yr})
= 8.10 \pm 0.05$, where the uncertainty is a combination of the
uncertainty in the exact vertical level of the fading line, and the
uncertainty regarding exactly when cluster disruption is significantly
more important than fading. In order to assess the latter uncertainty,
we applied a linear regression to the data points for $\log({\rm
Age/yr}) \ge 8.2$ and 8.4, respectively, resulting in the dotted and
dash-dotted fits in Fig.  \ref{agemasshist.fig}a. We note that for
these fits we have omitted the data points in the LMC's age gap,
between 3 and 13 Gyr (see Section \ref{cfr.sec}), since these would
obviously violate the underlying assumption of a constant CFR.

In addition, we need the slope of the fading line ($-0.648$; from
stellar population synthesis modelling) and the exact mass dependence
of the disruption time-scale, $\gamma$, defined as $t_{\rm dis}
\propto M_{\rm cl}^\gamma$ (Boutloukos \& Lamers 2003), in order to
derive the characteristic disruption time-scale. Boutloukos \& Lamers
(2003) derived $\gamma = 0.62$ from observations of a small but
diverse sample of galaxies containing rich cluster systems. Using
these parameters as our input, we derive a characteristic cluster
disruption time-scale for a $10^4$ M$_\odot$ cluster of $\log(
t_4^{\rm dis}/{\rm yr} ) = 9.9 \pm 0.1$. This is in good agreement
with Boutloukos \& Lamers (2002), who found $\log( t_4^{\rm dis} /
{\rm yr} ) = 9.7 \pm 0.3$ for a smaller sample of 478 clusters within
5 kpc from the centre of the LMC, in the age range $7.8 \le \log({\rm
Age/yr}) \le 10.0$ (data from Hodge 1988), while our result is also
qualitatively consistent with H03 who noticed very little destruction
of clusters on the high-mass end. This is not unexpected, considering
the low-density environment in which these clusters are found
(cf. Lamers et al. 2005).

Even with recent improvements (e.g., Gieles et al. 2005) to the simple
model of Boutloukos \& Lamers (2003), the characteristic cluster
destruction time-scales resulting from a more sophisticated,
non-instantaneous destruction process are very similar to those based
on the simple method (see Lamers et al. 2005 for a detailed
comparison, in particular their table 1).

Finally, for completeness and a full assessment of the uncertainties
involved, we need to address the effects of possibly having
underestimated the ages of the oldest sample clusters. The ages of the
oldest LMC clusters are well-known from detailed {\sl Hubble Space
Telescope}-based studies (e.g., Olsen et al. 1998) to all have
essentially the same age of $\log({\rm Age/yr} ) \simeq
10.1$. Therefore we reanalyzed Fig. \ref{agemasshist.fig}b, now
assuming this uniformly old age for all clusters for which we
determined $\log({\rm Age/yr}) \ge 9.5$. Under this assumption, the
resulting crossing time requires adjustment by $\Delta \log(t_{\rm
cross}/{\rm yr}) \simeq -0.05$, leading to a lower limit to the
characteristic disruption time-scale of $\log(t_4^{\rm dis}/{\rm yr})
= 9.8 \pm 0.1$, which is still in very good agreement with the earlier
determination by Boutloukos \& Lamers (2002). 

Our analysis in this section depends on the key assumption that the
CFR has remained approximately constant for the entire lifetime of the
LMC star cluster system, with the exception of the time span covered
by the age gap. That this assumption is justified to first order is
supported by two additional lines of evidence: (i) the disruption
lines in Fig. \ref{agemasshist.fig}a are representative of the full
LMC cluster sample for ages of $\log({\rm Age/yr}) \gtrsim 8.1$,
whether or not we include the oldest clusters that may possibly have
formed at a different CFR (if we leave out the oldest clusters, the
main effect is that the slope of the disruption line becomes less well
constrained); (ii) the slope of the disruption line in either case
corresponds to a mass dependence of the LMC cluster disruption
time-scale of $t_{\rm dis} \propto M_{\rm cl}^{\gamma = 0.56 \pm
0.07}$, which is -- within the uncertainties -- well inside the
approximately universal proportionality found by Boutloukos \& Lamers
(2003) for a range of galaxies in the local Universe, $\gamma = 0.62
\pm 0.06$.

\subsection{The LMC cluster formation rate}
\label{cfr.sec}

\begin{figure}
\psfig{figure=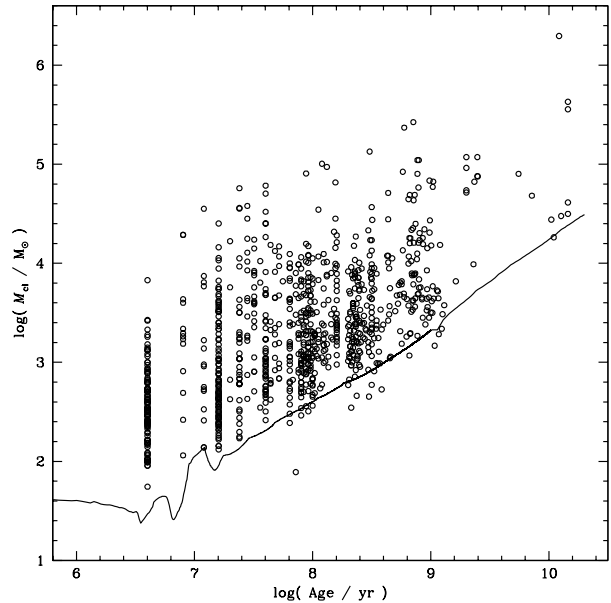,width=8.8cm}
\caption{\label{agemassdiag.fig}Distribution of the LMC clusters in
the (age vs. mass) plane. Overplotted is the expected detection limit
based on stellar population synthesis for a completeness limit of $M_V
= -3.5$ mag (H03), assuming no extinction. For a nominal extinction of
$A_V = 0.1$ mag (assuming the Calzetti attenuation law), the detection
limit will shift to higher masses by $\Delta \log( M_{\rm cl}/ M_\odot
) = 0.04$, which is well within the uncertainties associated with our
mass determinations. The features around 10 Myr are caused by the
appearance of red supergiants in the models.}
\end{figure}

In Fig. \ref{agemassdiag.fig} we show the distribution of the full LMC
cluster sample in the (age vs. mass) plane. For reasons of clarity, we
have not included error bars in this figure. Overplotted is the 50 per
cent completeness limit based on stellar population synthesis for
single-burst (``simple'') stellar populations (appropriate in the
context of star clusters), for an estimated $\sim 50$ per cent
completeness limit of $M_V = -3.5$ mag (H03), assuming a distance
modulus to the LMC of $(m-M)_0 = 18.48$ mag, and no extinction. For a
nominal extinction of $A_V = 0.1$ mag (assuming the Calzetti
extinction law), the equivalent detection limit will shift to higher
masses by $\Delta \log( M_{\rm cl}/ {\rm M}_\odot ) = 0.04$, which is
well within the uncertainties associated with our mass determinations.

For a constant CFR, we expect the number of data points to increase
gradually and evenly from young to old ages in the logarithmic
representation of Fig. \ref{agemassdiag.fig}. Any obviously clumped
subsets of data points may indicate either artifacts caused by our
computational routines, or real deviations from a constant CFR. The
effects of using interpolation of discrete isochrones for the
determinations of cluster ages, and the resulting artifacts in
age--mass space have been discussed extensively by Bastian et
al. (2005) and Gieles et al. (2005). Based on their discussions, two
features in Fig. \ref{agemassdiag.fig} in particular can be attributed
to artifacts caused by the fitting procedures: (i) the large density
of clusters at $\log( {\rm Age/yr} ) = 6.6$ is simply caused by the
fact that our youngest isochrone is at an age of 4 Myr (we are limited
by the age range spanned by the Padova isochrones on which the {\sc
galev} SSP models are based), and younger clusters would therefore be
assigned the minimum model age; (ii) the apparent overdensity around
$\log( {\rm Age/yr} ) = 7.2$ is caused by the discreteness of the
isochrones in this age range, where rapid changes occur in realistic
stellar populations; it was also noted by Gieles et al. (2005) and
Fall et al. (2005). Fortunately, however, from their age and mass
determinations of artificial cluster populations using very similar
procedures as done here, Gieles et al. (2005) find that they can
retrieve the ages and masses of, respectively, 87 and 97 per cent of
their input clusters to within 0.4 dex in $\log( {\rm Age/yr} )$ and
$\log( M_{\rm cl}/{\rm M}_\odot )$, respectively (see also Fall et
al. 2005). These retrieval rates are based on the assumption of a
constant CFR, and an initial power-law CMF with exponent $\alpha =
-2$.

In addition to these artifacts, we inspected two other apparent
overdensities in Fig. \ref{agemassdiag.fig}, at ($7.8 \le \log( {\rm
Age/yr} ) \le 8.0, 2.8 \le \log( M_{\rm cl}/{\rm M}_\odot ) \le 3.4$)
and $8.2 \le \log( {\rm Age/yr} ) \le 8.4$ (for the full mass range),
by displaying the locations of these cluster subsamples across the
LMC. Neither of these apparent overdensities are associated with
localised enhanced CFRs. In fact, any sufficiently large age and mass
range (i.e., covering at least twice the typical uncertainty in age
and mass, following the Nyquist sampling theorem) shows a fairly
homogeneous and relatively continuous (in terms of the cluster ages)
distribution of star clusters across the disk of the LMC. We note, as
a caveat, that the H03 cluster sample does not cover the entire LMC
(e.g., it does not cover the dense LMC bar region), although Massey
(2002) and H03 argue that their fields are representative of the LMC
stellar (and cluster) population as a whole.

The last tidal encounter between the LMC and the Small Magellanic
Cloud (SMC) occurred most likely about 0.2 to 0.5 Gyr ago (e.g.,
Heller \& Rohlfs 1994; Gardiner \& Noguchi 1996). One rotation period
of the LMC corresponds to $\sim 250$ Myr (Grebel \& Brandner 1999), or
$\log(t^{\rm orb}_{\rm LMC}/{\rm yr}) \simeq 8.4$. This implies that
if an apparent feature in age--mass space at an age of $\log({\rm
Age/yr}) \simeq 7.9$ was triggered by the LMC-SMC encounter, but
corresponds to a spatial distribution of its clusters that is smooth
and homogeneous across the LMC disk (i.e., at all radii), rotational
mixing will not have had enough time for this to have happened, or at
best marginally so. While such a gravitational interaction could, in
principle, have triggered a galaxy-wide burst of cluster formation in
the LMC, we would still expect to see a clearly enhanced population of
clusters in a relatively narrow age range in such a case (cf. de Grijs
et al. 2003a,b for the equivalent diagnostics in M82). Lacking better
age resolution for a larger distinct cluster subpopulation, we cannot
conclude with sufficient certainty that the last encounter between the
LMC and the SMC was significantly strong to have triggered the
formation of a well-defined cluster population in the LMC as a result;
alternative triggering mechanisms may have been responsible for the
formation of the smoothly distributed clusters at young(er) ages.

The only real {\it under}density of data points is the well-known LMC
cluster age gap, between $\sim 3$ and 13 Gyr (e.g., Da Costa 1991;
Geisler et al. 1997; Rich et al. 2001; Piatti et al. 2002; Bekki et
al. 2004), which is well reproduced in our broad-band age
redeterminations. We note that possible other underdensities in
Fig. \ref{agemassdiag.fig} occur (i) close to the 50 per cent
completeness limit and (ii) for low masses, where we have to take
stochastic sampling effects of the stellar initial mass function (IMF)
into account; these effects act in the sense that a cluster of low
mass may be dominated by a small number of high(er)-mass, and
therefore high(er)-luminosity stars, which in turn will cause the
photometric age to be (somewhat) overestimated. The exact amount of
this depends sensitively on the cluster mass (see Section
\ref{cmf.sec} below) and age. In particular, stochastic effects will
be most pronounced in the age ranges where rapid evolutionary stages
show up in stellar models, such as the appearance of red supergiants
at ages around $8-12$ Myr (e.g., Girardi \& Bica 1993; Santos \&
Frogel 1997). Clusters that have reached ages of several $\times 10^7$
yr will be populated by a significant fraction of post-main sequence
stars, and suffer less from stochastic effects in optical passbands,
even if they are of low mass (e.g., Girardi \& Bica 1993; Santos \&
Frogel 1997). The appearance of thermally-pulsing asymptotic giant
branch stars around ages of $\sim 1$ Gyr can potentially give rise to
large stochastic IMF sampling effects. However, these are most
important at near-infrared wavelengths, and of lower importance at the
optical wavelengths used for the age and mass estimates in this
paper. Additional, yet circumstantial supporting evidence for this
statement is provided by the age and mass distribution shown in
Fig. \ref{agemassdiag.fig}, where we do not detect any significant
under or overdensity {\it in mass} of clusters at or close to ages of
$\log({\rm Age/yr}) \sim 9.0$, although such effects could -- to some
extent -- be masked by the relatively small number $(\lesssim 100)$ of
clusters in the logarithmic age interval [8.8,9.2].

An alternative method to portray the CFR is shown in Fig.
\ref{agemasshist.fig}a, where we display the number of clusters formed
per unit age range. If we omit the three data points in the age gap,
the remaining data points for $\log( {\rm Agr/yr} ) \gtrsim 8$ are
well approximated by a linear relationship. Since a linear
relationship is predicted if the CFR has been constant, this approach
shows the approximate validity of this assumption once again. If we
assume that the CFR in the cluster age gap has also been constant,
although at a (much) lower level (cf. H03), then we derive that the
ratio of the CFR before and after the cluster age gap to that in the
period between $\sim 3$ and 13 Gyr, is about $(5\pm1) : 1$. H03,
considering the upper cluster masses, and assuming that the mass of
the most massive cluster is determined by size-of-sample effects
(which is approximately valid; H03), derived an equivalent ratio of
$\sim 10$, although with large uncertainties. In view of the
independent methods used by H03 and us, based on entirely different
diagnostics, these results are in reasonable agreement.

Finally, H03 suggested that the oldest sample clusters represent a
separate population from the younger LMC clusters. They state that
lower-mass old clusters could have been observed but are not in
practice. However, our new age and mass determinations do not concur
with this result; Fig. \ref{agemasshist.fig} shows that our 50 per
cent completeness limit describes the lower mass limit of the entire
cluster sample up to the oldest ages very well. Similarly, the upper
mass limit of the oldest clusters is commensurate with the upper mass
limit expected from size-of-sample effects (cf. H03), and does not
require that the oldest clusters represent a separate population.
However, because of the small number of old clusters, we cannot draw
any firmer conclusions on this issue based on the information at hand.

\section{Implications for the cluster mass function as a function of age}
\label{cmf.sec}

In Fig. \ref{agemasshist.fig}b we show the LMC cluster mass
distribution in a way that allows us to assess the importance of
disruption processes for the sample, in the presence of an
age-dependent detection (completeness) limit (see Boutloukos \& Lamers
2003). As we showed in de Grijs et al. (2003d), the slope of the
distribution for the highest masses is in essence a projection of the
{\it initial} CMF, provided that we can prove that
(semi-instantaneous) disruption will not yet have had the time to act
on these masses.

In Table \ref{massslopes.tab} we list the derived slopes for mass
ranges from the minimum masses indicated to the highest masses present
in our sample. We note that, for $\log(M_{\rm cl}/{\rm M}_\odot)
\gtrsim 3.0$ the CMF slopes derived are very stable (except for the
highest mass ranges, where the small number of clusters affects the
results), resulting in an intial CMF slope of $\alpha = -1.85 \pm
0.05$ as our best determination. This is significantly (at the
$3\sigma$ level) smaller than the ``universal'' initial CMF slopes of
$\alpha = -2$ often found in interacting and starburst galaxies, and
used as the basis for theoretical models of the evolution of young
cluster populations (see, e.g., de Grijs et al. 2003d for a
review). We will discuss the implications of this result in the
context of our discussion of Fig. \ref{clfs2.fig} below. We note that
for $\log(M_{\rm cl}/{\rm M}_\odot)\ge 3.0$, $\log(t_{\rm dis}/{\rm
yr}) \ge 9.3$; in this age range, the clusters that would have been
affected by ongoing (semi-instantaneous) disruption have already faded
to below the adopted completeness limit, so that we conclude that here
we are indeed observing the {\it initial} CMF slope. We note that the
low density of the LMC field, while to some small extent effective in
tidal stripping and evaporation, does not lead to a significant
breakdown of our assumption of (almost) instantaneous disruption: most
of the evaporated and stripped stars will be of low mass, which hence
contribute negligibly to the integrated luminosities we use for our
analysis.

Using the value of $\alpha = -1.85$, and the fading parameter $\zeta =
0.648$ (where fading of a cluster's flux, $F_\lambda$, is defined as
$F_\lambda \sim t^{-\zeta}$) from stellar population synthesis
(Boutloukos \& Lamers 2003), the predicted power-law slope for the
low-mass range is $(1/\zeta)-|\alpha|$. This slope is shown as the
dash-dotted line in Fig. \ref{agemasshist.fig}b, and matches the
observed mass distribution remarkably well (we have only applied a
vertical shift to the slope in order to match the cluster numbers in
the distributions).

\begin{table}
\caption[ ]{\label{massslopes.tab}Mass range-dependent LMC cluster
mass function slopes, based on magnitude-limited sampling
(Fig. \ref{agemasshist.fig}b).}
\center{
{\scriptsize
\begin{tabular}{ccc}
\hline
\hline
\multicolumn{1}{c}{$\log( M_{\rm cl}/{\rm M}_{\odot} )_{\rm min}$} &
\multicolumn{2}{c}{CMF slope}\\
\cline{2-3}
& \multicolumn{1}{c}{${\rm d} N_{\rm cl} /$} & ${\rm d} \log( N_{\rm
cl} ) /$ \\
& \multicolumn{1}{c}{${\rm d} (M_{\rm cl}/{\rm M}_\odot)$} &
\multicolumn{1}{c}{${\rm d} \log(M_{\rm cl}/{\rm M}_\odot)$} \\
\hline
2.25 & $-1.59 \pm 0.08$ & $-0.59 \pm 0.08$ \\
2.50 & $-1.67 \pm 0.06$ & $-0.67 \pm 0.06$ \\
2.75 & $-1.73 \pm 0.06$ & $-0.73 \pm 0.06$ \\
3.00 & $-1.85 \pm 0.05$ & $-0.85 \pm 0.05$ \\
3.25 & $-1.83 \pm 0.05$ & $-0.83 \pm 0.05$ \\
3.50 & $-1.85 \pm 0.06$ & $-0.85 \pm 0.06$ \\
3.75 & $-1.86 \pm 0.08$ & $-0.86 \pm 0.08$ \\
4.00 & $-1.84 \pm 0.09$ & $-0.84 \pm 0.09$ \\
4.25 & $-1.81 \pm 0.12$ & $-0.81 \pm 0.12$ \\
4.50 & $-1.80 \pm 0.15$ & $-0.80 \pm 0.15$ \\
4.75 & $-1.78 \pm 0.21$ & $-0.78 \pm 0.21$ \\
5.00 & $-1.55 \pm 0.19$ & $-0.55 \pm 0.19$ \\
\hline
\end{tabular}
}}
\end{table}

\begin{figure*}
\hspace*{0.7cm}
\psfig{figure=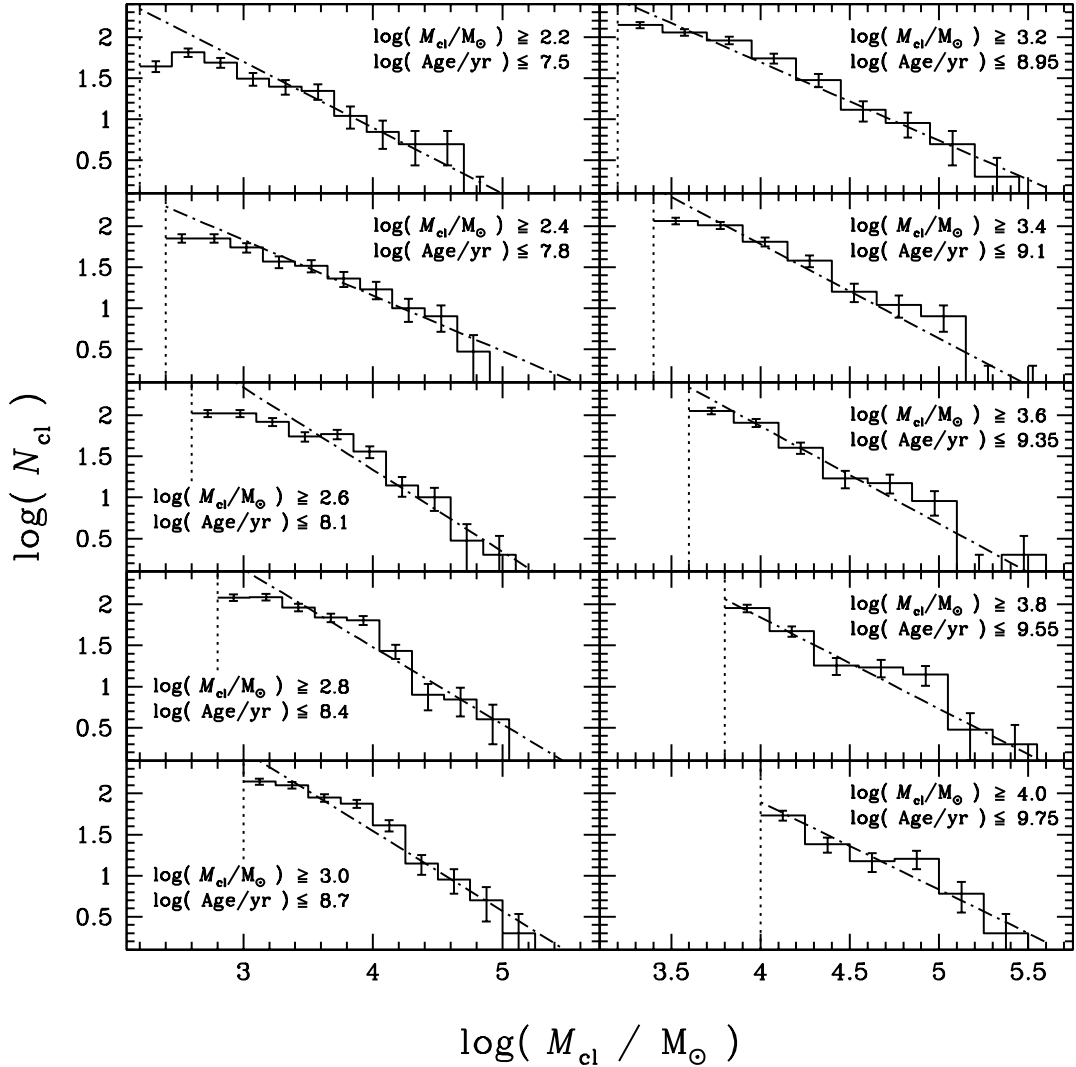,width=15cm}
\caption{\label{clfs2.fig}CMFs for statistically complete LMC cluster
subsamples. Age and mass ranges are indicated in the panel legends;
the vertical dotted lines indicate the lower mass limits
adopted. Error bars represent simple Poissonian errors, while the
best-fit lines were obtained for $\log( M_{\rm cl} / {\rm M}_\odot )
\ge 3.0$ for $\log( {\rm Age/yr} ) \le 8.7$, and for the full
available mass range for older maximum ages. Table \ref{mfslopes.tab}
summarises the best-fit parameters.}
\end{figure*}

Alternatively, we can assess the CMF and its possible evolution by
examining the CMFs of well-defined and well-understood subsamples
covering restricted mass ranges. Following Fall et al. (2005), in
Fig. \ref{clfs2.fig} we present mass-limited LMC cluster subsamples,
which are potentially physically more informative than
magnitude-limited subsamples. Mass-limited sample are less biased
toward young clusters than magnitude-limited samples. This is a novel
approach for the LMC star cluster system. In essence, a mass-limited
statistically complete sample of clusters implies an imposed age
limitation as well, so that for increasing masses, the upper age limit
increases, and so does the median age of the subsample. The mass and
age ranges used for the construction of the CMFs in
Fig. \ref{clfs2.fig} are listed in Table \ref{mfslopes.tab}, as well
as in the individual panels in the figure.

The error bars indicate the simple Poissonian uncertainties, and the
vertical dotted lines indicate the minimum mass limits adopted for a
particular cluster subsample; in all cases the corresponding age
limits are well below the expected disruption time-scales for even the
least massive clusters in a given sample. In Table \ref{mfslopes.tab}
we also include the CMF slopes we derive from each cluster subsample.
For samples up to $\log({\rm Age/yr}) \le 8.7$, we use a mass fitting
range of $\log(M_{\rm cl}/{\rm M}_\odot) \ge 3.0$; for the remaining
subsamples we use the full mass range available for the fits. We note
explicitly that using a lower age limit of $\log({\rm Age/yr}) = 6.8$
or 7.2, in order to avoid the artifact caused by the limited age
resolution of the isochrones used for the analysis does {\it not}
alter these results significantly (or may even strengthen these
results, albeit marginally).

The key result is that, for clusters with $\log(M_{\rm cl}/{\rm
M}_\odot) \ge 3.0$ {\it and} $\log({\rm Age/yr}) > 7.80$ (i.e., for
all but the two youngest subsamples), the CMF slopes are very close to
unity (see also H03 for similar results for the high-mass wings of
their age-limited CMFs\footnote{We note that H03 did not take the age
and mass-dependent completeness limits into account for the
construction of their CMFs (their fig. 6). However, a close comparison
between the mass fitting ranges they used on the one hand with Fig.
\ref{agemassdiag.fig} on the other reveals that they determined the
CMF slopes well inside the statistically complete mass range for each
age-limited cluster subsample. It is therefore not surprising, indeed
rather encouraging, that the results we obtain for a larger number of
subsamples for different age and mass ranges agree very well with
those of H03.}), which in this parameter space corresponds to an
initial CMF slope of $\alpha \simeq -2$ (see also the final column in
Table \ref{massslopes.tab} for a direct comparison). However, the
youngest two subsamples exhibit significantly shallower slopes for the
same mass range (see also Elmegreen \& Efremov 1997), while for
$\log(M_{\rm cl}/{\rm M}_\odot) < 3.0$ the observed CMFs turn down
even more significantly below the linear slopes indicated. We note
that the observed effect goes well beyond that expected from
uncertainties in the data's completeness limit: an uncertainty in the
completeness limit of 0.5 mag translates to an uncertainty in mass of
only $\Delta \log(M_{\rm cl}/{\rm M}_\odot) = 0.2$. As the Poissonian
error bars indicate, this flattening of the CMF slope is unlikely to
be caused by small-number statistics either, while cluster disruption
or variable external tidal effects are expected to have had a more
important effect on the low-mass extreme of {\it older} clusters than
on the younger clusters for which we notice this discrepancy. Thus,
this observation is contrary to theoretical predictions if the initial
CMF of the LMC cluster system were a power-law distribution modified
by cluster disruption processes. We recall that based on our
magnitude-limited sample analysis (Fig. \ref{agemasshist.fig}b), we
also concluded that the initial CMF slope appeared to be significantly
shallower, at the $3\sigma$ level, than the power-law slope, $\alpha =
-2$, expected for young star cluster systems.

Since we observe this behaviour towards the low-mass end for all mass
and age subsets of the full mass-limited cluster sample that include
masses $\log(M_{\rm cl}/{\rm M}_\odot) \lesssim 3$, as well as in our
magnitude-limited sample, random stochastic effects are unlikely to be
the primary cause (see also Girardi \& Bica 1993; Santos \& Frogel
1997). It is likely that this is a real effect (see also Elmegreen \&
Efremov 1997), and that the CMF slopes flatten significantly for
younger ages and lower-mass clusters in the LMC. Elmegreen \& Efremov
(1997) argue that the younger clusters are mostly unbound OB
associations (supporting Bica et al. 1996), of which some 90 per cent
will disperse by the time their constituent stars will reach an age of
$\sim 10^8$ yr. This is consistent with modern ideas on the formation
and dissolution of star clusters within the first $\sim 10^7$ yr of
their existence; most ($\sim 70-90$ per cent) of these newly formed
clusters will disperse on these time-scales, a process coined ``infant
mortality'' (e.g., Boily \& Kroupa 2003; Vesperini \& Zepf 2003;
Whitmore 2004; Bastian et al. 2005; Mengel et al. 2005; see also
Tremonti et al. 2001). If the process of infant mortality is mass
dependent (but see Whitmore 2004 for counterarguments), in the sense
that the lowest-mass clusters will dissolve preferentially, the result
will be a {\it flattening} of the CMF slopes with increasing mean age,
which is contrary to the apparent change of slope in
Fig. \ref{clfs2.fig}. Whether or not the youngest ($\lesssim 10$ Myr
old) clusters will dissolve because of this infant mortality scenario,
the significant differences among the LMC's CMF slopes as a function
of age are predominantly driven by the youngest subset of our cluster
sample.

Thus, these results may imply that the initial CMF slope of the {\it
combined} (i.e., both bound and unbound) LMC cluster system is {\it
not} well represented by a power-law, although we cannot disentangle
the unbound from the bound clusters at the youngest ages. In addition,
we recently presented observational evidence and theoretical arguments
against an initial power-law CMF in M82's intermediate-age starburst
region, M82 B (de Grijs, Parmentier \& Lamers 2005), and in favour of
an initial log-normal CMF in the Antennae interacting system, NGC
4038/39 (Anders et al. 2005). Our detailed analysis of the LMC cluster
mass distributions as a function of age and mass may therefore have
uncovered supporting new evidence that star clusters -- at least in
the low-density environment of the LMC -- may not form following a
power-law distribution that holds strength down to masses much below a
few $\times 10^3$ M$_\odot$, where the power-law fits to the
subsamples in Fig. \ref{clfs2.fig} break down.


\begin{table}
\caption[ ]{\label{mfslopes.tab}LMC cluster mass function slope as a
function of minimum mass and maximum age, based on mass-limited
samples (Fig. \ref{clfs2.fig}).}
\center{
{\scriptsize
\begin{tabular}{ccc}
\hline
\hline
\multicolumn{1}{c}{$\log( M_{\rm cl}/{\rm M}_{\odot} )_{\rm min}$} &
\multicolumn{1}{c}{$\log( {\rm Age/yr} )_{\rm max}$} & CMF slope \\
& & \multicolumn{1}{c}{[${\rm d} \log( N_{\rm cl} ) / {\rm d} \log(
M_{\rm cl}/{\rm M}_\odot)$]} \\
\hline
2.2 & 7.50 & $-0.80 \pm 0.10$ \\
2.4 & 7.80 & $-0.68 \pm 0.06$ \\
2.6 & 8.10 & $-1.00 \pm 0.10$ \\
2.8 & 8.40 & $-0.94 \pm 0.10$ \\
3.0 & 8.70 & $-0.98 \pm 0.08$ \\
3.2 & 8.95 & $-0.95 \pm 0.06$ \\
3.4 & 9.10 & $-1.15 \pm 0.11$ \\
3.6 & 9.35 & $-1.18 \pm 0.14$ \\
3.8 & 9.55 & $-1.10 \pm 0.12$ \\
4.0 & 9.75 & $-1.06 \pm 0.15$ \\
\hline
\end{tabular}
}}
\end{table}

\section{Summary and conclusions}
\label{summary.sec}

By combining integrated properties with resolved stellar population
studies, the LMC cluster system offers the unique chance to
independently check the accuracy of age (and corresponding mass)
determinations based on broad-band SEDs. In this paper, we have
reanalyzed the broad-band LMC cluster SEDs based on the data of Massey
(2002) and H03, using a newly developed SED analysis approach. We
compare our new age determinations with (i) those of H03 using the
same data set but a different approach, and (ii) those of
Pietrzy\'nski \& Udalski (2000) using CMD fitting, in order to set the
tightest limits yet on the accuracy of (absolute) age determinations
based on broad-band SEDs, and therefore on the usefulness of such an
approach.

We note a significant systematic effect between the age differences of
H03 on the one hand, and those of both the OGLE-II team and our own
redeterminations on the other. It appears that these systematic
differences are caused by H03's conversions of the photometry to a
different filter system. We emphasize and warn that the {\it actual}
filter systems used for the observations should be used for the most
accurate parameter analysis, instead of using filter conversion
equations, in order to achieve more accurate derivations of the
cluster ages and the corresponding masses.

Based on this comparison, and additionally on a detailed assessment of
the age-metallicity and age-extinction degeneracies, we conclude that
our broad-band SED fits yield reliable ages, with statistical {\it
absolute} uncertainties within $\Delta\log( \mbox{Age/yr}) \simeq 0.4$
overall. Thus, in addition to our conclusion in de Grijs et al.
(2005) that we can retrieve prominent features in the cluster age
distribution to within $\Delta \langle \log( {\rm Age / yr} ) \rangle
\le 0.35$ using a variety of approaches based on broad-band SEDs
modelling, we have now also shown that the associated {\it intrinsic}
statistical uncertainties involved in cluster age determinations are
of a very similar magnitude.

The LMC's CFR has been roughly constant outside of the well-known age
gap between $\sim 3$ and 13 Gyr, when the CFR was a factor of $\sim 5$
lower (assuming a roughly constant rate during the entire
period). There are no clear observational signatures of an enhanced
CFR associated with the last tidal encounter between the LMC and the
SMC, while we argue that the combination of the relevant time-scales,
i.e., the LMC's rotation period and the time since the last LMC-SMC
encouter, has been insufficient to wash out any such signatures, if
they had been present. An alternative triggering mechanism for the
young(er) clusters may be needed.

Using a simple approach to derive the characteristic cluster
disruption time-scale, we find that $\log(t_4^{\rm dis}/{\rm yr}) =
9.9 \pm 0.1$, where $t_{\rm dis} = t_4^{\rm dis} (M_{\rm cl}/10^4 {\rm
M}_\odot)^{0.62}$, for the LMC cluster system. This is consistent with
earlier, preliminary work for a smaller cluster sample. This long
characteristic disruption time-scale implies that hardly any of our
LMC sample clusters are affected by significant disruptive processes,
so that we are in fact observing the {\it initial} CMF. Using a
variety of complementary techniques, we conclude that the older
cluster (sub)samples show CMF slopes that are fully consistent with
the $\alpha \simeq -2$ slopes generally observed in young star cluster
systems. The youngest clusters in our sample show shallower slopes, at
least below masses of a few $\times 10^3$ M$_\odot$, and possibly
evidence for a turn-over. This is contrary to dynamical expectations
and may imply that the initial CMF slope of the LMC cluster system as
a whole is {\it not} well represented by a power-law down to the
lowest masses, although we cannot disentangle the unbound from the
bound clusters at the youngest ages.

\section*{acknowledgments} We thank Deirdre Hunter and Bruce Elmegreen
for kindly providing us with their LMC cluster photometry, Phil Massey
for helpful discussions regarding the filter sets used, Jon Holtzman
for sending us the unpublished Landolt KPNO filter transmission
curves, and Uta Fritze--v. Alvensleben and Henny Lamers for useful
comments. We acknowledge helpful comments on the OGLE-II cluster data
from Andrzej Udalski and Grzegorz Pietrzy\'nski. We also acknowledge
research support from and hospitality at the International Space
Science Institute in Berne (Switzerland), as part of an International
Team programme. RdG is grateful for hospitality at the National
Astronomical Observatories of the Chinese Academy of Sciences in
Beijing -- and in particular to Deng Li-Cai -- where most of this work
was completed; travel support was provided by the Royal Society under
the UK's Office of Science and Technology's UK-China science network
scheme, for one-to-one visits. This research has made use of NASA's
Astrophysics Data System Abstract Service.


\begin{thebibliography}{}

\bibitem[]{} Anders P., Bissantz N., Fritze--v. Alvensleben U., de
Grijs R., 2004, MNRAS, 347, 196

\bibitem[]{} Anders P., Fritze--v. Alvensleben U., 2003, A\&A, 401,
1063

\bibitem[]{} Anders P., Bissantz N., Boysen L., de Grijs R.,
Fritze--v. Alvensleben U., 2005, Nature Physics, submitted

\bibitem[]{} Barbaro G., Olivi F.M., 1991, AJ, 101, 922

\bibitem[]{} Bastian N., Gieles M., Lamers H.J.G.L.M., Scheepmaker
R.A., de Grijs R., 2005, A\&A, 431, 905

\bibitem[]{} Beasley M.A., Hoyle F., Sharples R.M., 2002, MNRAS, 336,
168

\bibitem[]{} Bekki K., Couch W.J., Beasley M.A., Forbes D.A., Chiba
M., Da Costa G.S., 2004, ApJ, 610, L93

\bibitem[]{} Bica E., Alloin D., Santos J.F.C. Jr., 1990, A\&A, 235,
103

\bibitem[]{} Bica E., Dottori H., Pastoriza M., 1986, A\&A, 156, 261

\bibitem[]{} Boily C.M., Kroupa P., 2003, MNRAS, 338, 665

\bibitem[]{} Boutloukos S.G., Lamers H.J.G.L.M., 2003, MNRAS, 338, 717

\bibitem[]{} Boutloukos S.G., Lamers H.J.G.L.M., 2002, in:
Extragalactic Star Clusters, IAU Symp. 207, Geisler D., Grebel E.K.,
Minniti D., (San Francisco: ASP), p.703

\bibitem[]{} Calzetti D., 1997, AJ, 113, 162

\bibitem[]{} Calzetti D., Armus L., Bohlin R.C., Kinney A.L.,
Koornneef J., Storchi-Bergmann T., 2000, ApJ, 533, 682

\bibitem[]{} Calzetti D., 2001, PASP, 113, 1449

\bibitem[]{} Cardelli J.A., Clayton G.C., Mathis J.S., 1989, ApJ, 345,
245

\bibitem[]{} Charlot S., Bruzual A.G., 1991, ApJ, 367, 126

\bibitem[]{} Chiosi C., Bertelli G., Bressan A., 1988, A\&A, 196, 84

\bibitem[]{} Da Costa G.S., 1991, in IAU Symp. 148, The Magellanic
Clouds, Haynes R., Milne D., eds., (Dordrecht: Kluwer), p. 183

\bibitem[]{} de Grijs R., O'Connell R.W., Gallagher J.S., 2001, AJ, 121,
768

\bibitem[]{} de Grijs R., Bastian N., Lamers H.J.G.L.M., 2003a, MNRAS,
340, 197

\bibitem[]{} de Grijs R., Bastian N., Lamers H.J.G.L.M., 2003b, ApJ,
583, L17

\bibitem[]{} de Grijs R., Fritze-v. Alvensleben U., Anders P.,
Gallagher J.S.  {\sc iii}, Bastian N., Taylor V.A., Windhorst R.A.,
2003c, MNRAS, 342, 259

\bibitem[]{} de Grijs R., Anders P., Lynds R., Bastian N., Lamers
H.J.G.L.M., O'Neill E.J., Jr., 2003d, MNRAS, 343, 1285

\bibitem[]{} de Grijs R., Anders P., Lamers H.J.G.L.M., Bastian N.,
Parmentier G., Sharina M.E., Yi S., 2005, MNRAS, 359, 874

\bibitem[]{} de Grijs R., Parmentier G., Lamers H.J.G.L.M., 2005,
MNRAS, in press (astro-ph/0509721; doi:
10.1111/j.1365-2966.2005.09640.x)

\bibitem[]{} Elson R.A.W., Fall S.M., 1988, AJ, 96, 1383

\bibitem[]{} Elmegreen B.G., Efremov Y.N., 1997, ApJ, 480, 235

\bibitem[]{} Fall S.M., Chandar R., Whitmore B.C., 2005, ApJ, 631,
L133

\bibitem[]{} Fritze-v.  Alvensleben U., 1998, A\&A, 336, 83

\bibitem[]{} Fritze-v.  Alvensleben U., 1999, A\&A, 342, L25

\bibitem[]{} Gardiner L.T., Noguchi M., 1996, MNRAS, 278, 191

\bibitem[]{} Geisler D., Bica E., Dottori H., Clari\'a J.J.,
Piatti A.E., Santos J.F.C., Jr., 1997, AJ, 114, 1920

\bibitem[]{} Gieles M., Bastian N., Lamers H.J.G.L.M., Mout J.N.,
2005, A\&A, 441, 949

\bibitem[]{} Girardi L., Bica E., 1993, A\&A, 274, 279

\bibitem[]{} Grebel E.K., Brandner W., 1999, in: New Views of the
Magellanic Clouds, IAU Symp. 190, Chu Y.-H., Suntzeff N., Hesser J.,
Bohlender D., (San Francisco: ASP), p. 470

\bibitem[]{} Heller P., Rohlfs K., 1994, A\&A, 291, 743

\bibitem[]{} Hodge P., 1988, PASP, 100, 576

\bibitem[]{} Holtzman J.A., Burrows C.J., Casertano S., Hester J.J.,
Trauger J.T., Watson A.M., Worthey G., 1995, PASP, 107, 1065

\bibitem[]{} Hunter D.A., Elmegreen B.G., Dupuy T.J., Mortonson M.,
2003, AJ, 126, 1836 (H03)

\bibitem[]{} Kurth O.M., Fritze-v. Alvensleben U., Fricke K.J., 1999,
A\&AS, 138, 19

\bibitem[]{} Lamers H.J.G.L.M., Gieles M., Portegies Zwart S., 2005,
A\&A, 429, 173

\bibitem[]{} Landolt A.U., 1992, AJ, 104, 340

\bibitem[]{} Leitherer C., et al., 1999, ApJS, 123, 3 (Starburst99)

\bibitem[]{} Leitherer C., Li I.-H., Calzetti D., Heckman T.M., 2002,
ApJS, 140, 303

\bibitem[]{} Massey P., 2002, ApJS, 141, 81

\bibitem[]{} Mengel S., Lehnert M.D., Thatte N., Genzel R., 2005,
A\&A, 443, 41

\bibitem[]{} Meurer G.R., Cacciari C., Freeman K.C., 1990, AJ, 99,
1124

\bibitem[]{} Meurer G.R., Heckman T.M., Leitherer C., Kinney A., Robert
C., Garnett D.R., 1995, AJ, 110, 2665

\bibitem[]{} Miller B.W., Whitmore B.C., Schweizer F., Fall S.M., 1997,
AJ, 114, 2381

\bibitem[]{} Olsen K.A.G., Hodge P.W., Mateo M., Olszewski E.W.,
Schommer R.A., Suntzeff N.B., Walker A.R., 1998, MNRAS, 300, 665

\bibitem[]{} Olszewski E.W., Schommer R.A., Suntzeff N.B., Harris
H.C., 1991, AJ, 101, 515

\bibitem[]{} Piatti A., Sarajedini A., Geisler D., Bica E., Clari\'a
J.J., 2002, MNRAS, 329, 556

\bibitem[]{} Pietrzy\'nski G., Udalski A., 2000, AcA, 50, 337
(OGLE-II)

\bibitem[]{} Reed B.C., 1985, PASP, 97, 120

\bibitem[]{} Rich R.M., Shara M.M., Zurek D., 2001, AJ, 122, 842

\bibitem[]{} Santos J.F.C., Jr., Bica E., Clar\'\i a J.J., Piatti
A.E., Girardi L.A., Dottori H., 1995, MNRAS, 276, 1155

\bibitem[]{} Santos J.F.C., Jr., Frogel J.A., 1997, ApJ, 479, 764

\bibitem[]{} Santos J.F.C., Jr., Piatti A.E., 2004, A\&A, 428, 79

\bibitem[]{} Schlegel D.J., Finkbeiner D.P., Davis M., 1998, ApJ, 500,
525

\bibitem[]{} Schulz J., Fritze-v. Alvensleben U., M\"oller C.S., Fricke
K.J., 2002, A\&A, 392, 1

\bibitem[]{} Searle L., Sargent W.L.W., Bagnuolo W.G., 1973, ApJ, 179,
427

\bibitem[]{} Searle L., Wilkinson A., Bagnuolo W.G., 1980, ApJ, 239,
803 (SWB)

\bibitem[]{} Tremonti C.A., Calzetti D., Leitherer C., Heckman T.M.,
2001, ApJ, 555, 322

\bibitem[]{} Udalski A., Kubiak M., Szyma\'nski M., 1997, AcA, 47, 319

\bibitem[]{} Udalski A., Soszy\'nski I., Szyma\'nski M., Kubiak M.,
Pietrzy\'nski G., Wo\'zniak P., $\dot{\rm Z}$ebru\'n K., 1999, AcA, 49, 223

\bibitem[]{} Vallenari A., Bettoni D., Chiosi C., 1998, A\&A, 331, 506

\bibitem[]{} Vesperini E., Zepf S.E., 2003, ApJ, 587, L97

\bibitem[]{} Whitmore B.C., 2004, in: The Formation and Evolution of
Massive Young Star Clusters, ASP Conf. Ser., vol. 322, Lamers
H.J.G.L.M., Smith L.J., Nota A., eds., (ASP: San Francisco), p. 419

\end{thebibliography}
\end{document}